\documentclass[pre,twocolumn,showpacs]{revtex4}

\usepackage{graphicx,amsmath}

\newcommand{\figwidth}{0.9\columnwidth}
\newcommand{\eq}[1]{Eq.(\ref{#1})}
\newcommand{\fig}[1]{Fig.~\ref{#1}}

\newcommand{\olcite}[1]{Ref.~\onlinecite{#1}}
\newcommand{\sect}[1]{Section~\ref{#1}}

\newcommand{\kb}{k_{\rm B}}
\newcommand{\et}{\epsilon^\star}
\newcommand{\mt}{\mu^\star}
\newcommand{\gt}{\gamma(\theta)}

\begin{document}

\title{Anchoring effects at the isotropic-nematic interface in liquid 
crystals}

\author{R. L. C. Vink}
\affiliation{Institut f\"ur Theoretische Physik II, Heinrich Heine
Universit\"at D\"usseldorf, Universit\"atsstra{\ss}e 1, 40225
D\"usseldorf, Germany}

\date{\today}

\begin{abstract} The isotropic-to-nematic transition in liquid crystals is 
studied in $d=3$ spatial dimensions. A simulation method is proposed to 
measure the angle dependent interfacial tension $\gt$, with $\theta$ the 
anchoring angle of the nematic phase at the interface. In addition, an 
alternative liquid crystal model is introduced, defined on a lattice. The 
advantage of the lattice model is that accurate simulations of anchoring 
effects become possible. For the lattice model, $\gt$ depends sensitively 
on the nearest-neighbor pair interaction, and both stable and metastable 
anchoring angles can be detected. We also measure $\gt$ for an {\it 
off-lattice} fluid of soft rods. For soft rods, only one stable anchoring 
angle is found, corresponding to homogeneous alignment of the nematic 
director in the plane of the interface. This finding is in agreement with 
most theoretical predictions obtained for hard rods. \end{abstract}

%% 83.80.Xz 	Liquid crystals: nematic, cholesteric, smectic, discotic
%% 68.05.-n 	Liquid-liquid interfaces
%% 68.03.Cd 	Surface tension and related phenomena
%% 64.70.Md 	Transitions in liquid crystals
%% 61.30.Hn 	Surface phenomena: alignment, anchoring


\maketitle

\section{Introduction}

A fluid consisting of elongated molecules is more difficult to describe 
than one in which the molecules are simply spherical. In the case of 
elongated molecules, there are not only translational degrees of freedom, 
but also orientational ones. This additional complexity gives rise to many 
interesting effects, not found in spheres. For example, infinitely slender 
rods in three dimensions undergo a first-order phase transition from an 
isotropic to a nematic phase, provided the density is sufficiently high 
\cite{onsager:1949}. Both the isotropic and the nematic phase lack 
translational order, but in the nematic phase the rods have aligned, 
giving rise to long-range orientational order.

\begin{figure}
\begin{center}
\includegraphics[clip=,width=\columnwidth]{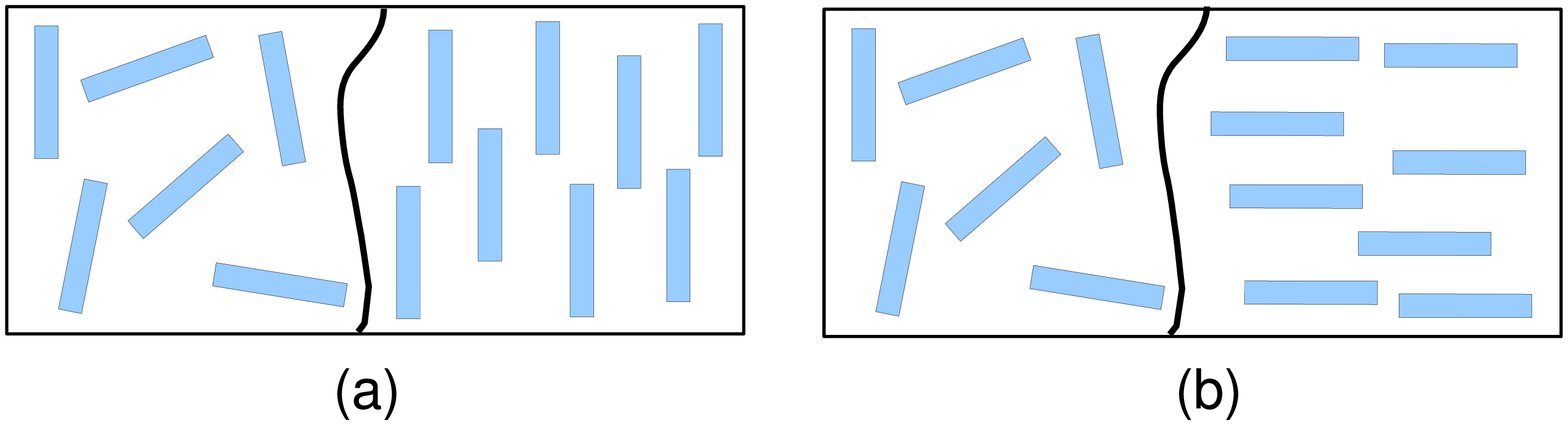}
\caption{\label{problem} Schematic representation of isotropic-nematic 
phase coexistence. The isotropic phase is on the left, the nematic on the 
right. Shown are (a) homogeneous anchoring, and (b) homeotropic 
anchoring.}
\end{center}
\end{figure}

The orientation of the nematic phase is an important quantity. In 
applications involving nematics at walls, the angle of the nematic 
director at the wall is often crucial. This angle is called the tilt or 
anchoring angle. Typically, there is a preferred tilt angle the nematic 
phase will assume, but the precise value depends sensitively on factors 
such as surface chemistry, particle shape, and temperature 
\cite{jerome:1991, patel.yokoyama:1993, barmes:021705}. Similarly, 
anchoring effects also occur at the isotropic-to-nematic (IN) transition. 
The first-order nature of that transition implies phase coexistence, 
whereby isotropic domains coexist with nematic domains, separated by 
interfaces. As \fig{problem} shows, the orientation of the nematic phase 
with respect to the interface becomes an additional parameter. In 
\fig{problem}(a), the nematic director points in the plane of the 
interface, which is called planar or homogeneous alignment. In 
\fig{problem}(b), the director is perpendicular to the interface, which is 
known as homeotropic alignment.

From symmetry considerations alone, it is clear that homogeneous and 
homeotropic anchoring are different. For homeotropic anchoring, there is 
still rotational symmetry around the interface normal; for homogeneous 
anchoring, no such symmetry is present. This difference is known to affect 
the spectrum of capillary waves. For homogeneous anchoring, the spectrum 
becomes anisotropic in the short wavelength limit 
\cite{akino.schmid.ea:2001, elgeti.schmid:2005, schmid.germano.ea:2007, 
wolfsheimer:061703}. In contrast, for homeotropic anchoring, the spectrum 
remains isotropic at all wavelengths. In other words, as this example 
shows, the anchoring angle affects the interfacial properties 
qualitatively. Given a set of particle interactions, it is therefore 
important to be able to predict the anchoring angle. This has lead to the 
concept of an angle dependent interfacial tension $\gt$, with $\theta$ the 
tilt or anchoring angle. Here, $\theta$ is defined as the angle between 
the nematic director and the plane of the IN interface. Homogeneous 
($\theta=0$) and homeotropic ($\theta=90$) anchoring are most common, 
although $\theta$ could, in principle, be anywhere between 0 and 90 
degrees. In theoretical investigations, the anchoring angle is given by 
the angle which minimizes $\gt$. For hard rods, this is typically 
$\theta=0$, corresponding to homogeneous anchoring 
\cite{chen.noolandi:1992, velasco.mederos:2002, mcmullen:1988, 
physreva.42.6042, koch.harlen:1999, shundyak.roij:2001}, but the precise 
behavior is quite subtle. For example, the results of 
\olcite{chen.noolandi:1992} also suggest that homeotropic anchoring may be 
metastable. In addition, for very short rods, anchoring angles between 0 
and 90 degrees have also been reported \cite{physreva.42.6042}.

Unfortunately, it remains difficult to verify these theoretical findings 
in a computer simulation. On the one hand, efficient simulation 
methodology for problems of this kind is scarce. The state-of-the-art is 
to extract $\gt$ from the anisotropy of the pressure tensor 
\cite{allen:2000*b, mcdonald:2000, akino.schmid.ea:2001}, a technique 
which is somewhat prone to statistical error. On the other hand, the 
particle interactions used in many theoretical investigations are not 
convenient for simulations. The hard rod potential, for instance, often 
used in theory, gives rise to a very small interfacial tension. In order 
to stabilize the IN interface, simulations of hard rods require huge 
system sizes, implying long equilibration times and, consequently, data 
with considerable statistical uncertainty.

The purpose of this paper is to improve on this state of affairs. The 
primary aim is to present a simulation method capable of measuring the 
angle dependent interfacial tension accurately. The method is presented in 
\sect{sec:sim}. As it turns out, the method is general, and applies to 
lyotropic (density driven) systems, such as rods or platelets, as well as 
to thermotropic (temperature driven) lattice systems. The second aim is to 
introduce a new liquid crystal model, one which is easy to simulate, but 
which nevertheless features an IN transition with anchoring effects. The 
model we propose is defined on a {\it lattice}, and resembles the 
Lebwohl-Lasher (LL) model \cite{physreva.6.426}, but with two essential 
modifications. Since the model is easy to simulate, it lends itself 
perfectly for an investigation of anchoring effects. The liquid crystal 
model, and the subsequent determination of its $\gt$, are presented in 
\sect{res1}. Next, in \sect{res2}, we determine $\gt$ for a fluid of soft 
rods. These particles are already more complicated to simulate. 
Nevertheless, guided by the experience obtained for the simple lattice 
model, a meaningful interpretation of the simulation data is possible. We 
end with a summary and outlook in the last section.

\section{Simulation Method}
\label{sec:sim}

In this Section, we present our method to extract the angle dependent 
interfacial tension $\gt$ in liquid crystals. The use of 
so-called {\it order parameters} is crucial for our method: suitable order 
parameters are therefore discussed first. Next, we show how the order 
parameter distribution may be used to obtain phase coexistence properties, 
as well as interfacial tensions, which summarizes the key ingredients of 
previous work \cite{vink.schilling:2005, vink.wolfsheimer.ea:2005}. 
Finally, we show how this methodology can be modified, to also capture the 
angular dependence of the interfacial tension, by means of a simple 
constraint.

\subsection{Order parameters}
\label{op}

Since we are dealing with the IN transition in liquid crystals, a suitable 
order parameter is the nematic order parameter $S$, defined as the maximum 
eigenvalue of the orientational tensor $Q$, whose elements read as:
\begin{equation}\label{eq:s}
Q_{\alpha\beta} = \frac{1}{2 N} \sum_{i=1}^N
   \left( 3 d_{i\alpha} d_{i\beta} - \delta_{\alpha\beta} \right).
\end{equation}
Here, $d_{i\alpha}$ is the $\alpha$ component ($\alpha = x,y,z$) of the 
orientation $\vec{d}_i$ of molecule $i$ (normalized to unity), 
$\delta_{\alpha\beta}$ is the Kronecker delta, and $N$ the number of 
molecules. Note that $S$ is invariant under the inversion $\vec{d}_i \to 
-\vec{d}_i$ of single molecules, which is the characteristic symmetry of 
liquid crystals, and also that $S$ does not depend on the center of mass 
coordinates. In the isotropic phase, $S$ is close to zero. In the nematic 
phase, where the molecules have aligned, $S$ is close to unity. Another 
important quantity is the (normalized) eigenvector $\vec{n}=(n_x,n_y,n_z)$ 
associated with $S$. The vector $\vec{n}$ is called the director, and it 
corresponds to the overall preferred direction of the molecular 
orientations in the nematic phase. Again, the directions $\vec{n} \to 
-\vec{n}$ are equivalent: the convention in this work is to pick the 
vector with $n_z>0$.

The nematic order parameter, being zero in the isotropic phase and (close 
to) unity in the nematic phase, is a convenient quantity to detect the IN 
transition in liquid crystals. However, different quantities may be used 
as well. For example, in thermotropic (temperature driven) liquid 
crystals, there is also an energy difference between the isotropic (high 
energy) and nematic (low energy) phase. Therefore, in thermotropic 
systems, energy may also be used as order parameter. Similarly, in 
lyotropic (density driven) liquid crystals, such as studied by Onsager 
\cite{onsager:1949}, there is also a density difference between the 
isotropic (low density) and nematic (high density) phase. Therefore, in 
lyotropic systems, density is also a valid order parameter.

\subsection{Order parameter distributions}
\label{opd}

\begin{figure}
\begin{center}
\includegraphics[clip=,width=\figwidth]{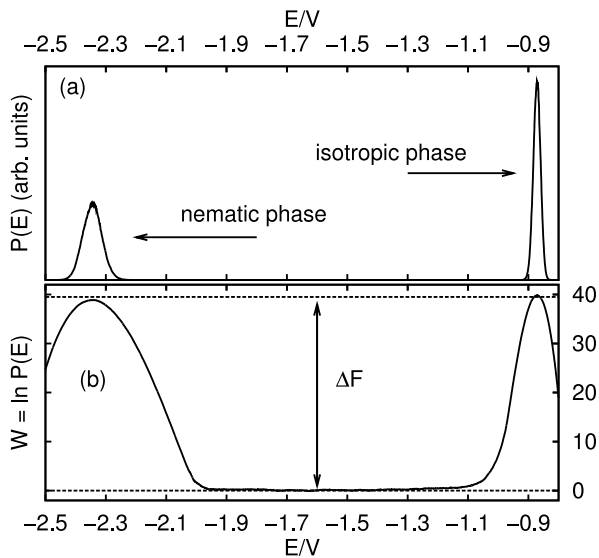}

\caption{\label{bimodal} (a) Coexistence distribution $P(E)$ of a 
thermotropic liquid crystal, interacting via \eq{eq:ll} with $p=10$ and 
$\nu=0.5$, at the transition (inverse) temperature $\et \approx 
1.188$. The simulation box dimensions were $L=15$ and $D=40$. (b) The 
logarithm of the same distribution.}

\end{center}
\end{figure}

Our method to obtain $\gt$ is based on the order parameter 
distribution $P(X)$, defined as the probability to observe the order 
parameter $X$ during the simulation. For liquid crystals, suitable choices 
for $X$ were given above. As is well known, at a first-order phase 
transition, the distribution $P(X)$ becomes double-peaked (bimodal). An 
example is provided in \fig{bimodal}(a), which shows the energy 
distribution $P(E)$ of a thermotropic liquid crystal (details are provided 
in \sect{res1}). In thermotropic systems, $P(E)$ becomes bimodal at the 
transition temperature. The precise value is determined using the 
``equal-area'' rule \cite{binder.landau:1984}, whereby the temperature is 
tuned such that the area under both peaks is equal. Of course, in a 
lyotropic system, one would need to tune the chemical potential.

\begin{figure}
\begin{center}
\includegraphics[clip=,width=0.6\columnwidth]{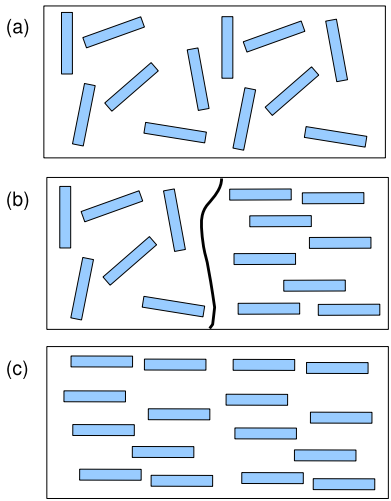}

\caption{\label{snap} Schematic simulation snapshots in (a) the bulk 
isotropic phase, (b) the coexistence region, and (c) the bulk nematic 
phase.}

\end{center}
\end{figure}

From the bimodal energy distribution of \fig{bimodal}(a), bulk properties 
can readily be extracted. The peak at high energy, for example, yields the 
energy density of the isotropic phase; the peak at low energy of the 
nematic phase. Even more information is contained in the logarithm $W = 
\ln P(X)$, see \fig{bimodal}(b). Note that $W$ corresponds to {\it minus} 
the free energy of the system. We now observe a distinct flat region 
between the peaks. The origin of this flat region can be understood from 
simulation snapshots, shown schematically in \fig{snap}. When the system 
is in the high-energy peak, simulation snapshots reveal a homogeneous 
isotropic phase (a). In the low-energy peak, snapshots reveal a 
homogeneous nematic phase (c). At intermediate energy, coexistence between 
an isotropic and nematic domain is revealed, separated by an interface 
(b). Note that, due to periodic boundary conditions, two such interfaces 
are actually present. Provided the simulation box is large enough so as to 
accommodate two non-interacting interfaces, the order parameter can be 
varied (over a limited range) with no cost in the free energy at all, and 
hence a flat region in $W$.

The presence of a flat region in $W$ naturally allows for an estimate of 
the interfacial tension \cite{binder:1982}. In these cases, the height of 
the free energy barrier $\Delta F$ in \fig{bimodal}(b) may be associated 
with the free energy cost of having two interfaces in the system. Since 
the interfacial tension is defined as the excess free energy per unit 
area, one simply has $\gamma = \Delta F / 2A$, with $A$ the area of one 
interface. It was later recognized that \cite{grossmann.laursen:1993}, in 
an elongated $L \times L \times D$ simulation box, with $D \gg L$, the 
interfaces form perpendicular to the elongated direction, since this 
minimizes the total amount of interface in the system. This leads to 
$A=L^2$, and consequently
\begin{equation}\label{eq:st}
  \gamma = \Delta F / 2L^2.
\end{equation}
In previous work, the above ideas were successfully applied to the IN 
transition in fluids of rods \cite{vink.schilling:2005, 
vink.wolfsheimer.ea:2005} and platelets \cite{beek:087801}. Implementation 
details are also provided in these references. Of particular importance is 
the use of a biased sampling scheme \cite{virnau.muller:2004}, such that 
the simulation frequently traverses between the isotropic and the nematic 
phase.

\subsection{Measuring $\gt$}

Next, we describe how to modify the above methodology to also extract the 
angular dependence of the interfacial tension. We again use an elongated 
simulation box with periodic boundary conditions. The box is spanned by 
the vectors $L \hat{x}$, $L \hat{y}$, and $D \hat{z}$, with $D >> L$. As 
usual, $\hat{x} = (1,0,0)$, $\hat{y}=(0,1,0)$, and $\hat{z}=(0,0,1)$ 
denote standard Cartesian unit vectors. The key additional ingredient is 
to add a constraint to the Hamiltonian, such that the total energy of the 
system becomes $E = E_0 + E_c$. Here, $E_0$ is the energy of the 
unconstrained system. For example, in a thermotropic system, $E_0$ could 
be the LL potential. In a lyotropic system, it could be the 
potential of hard rods. The constraint energy $E_c$ should 
fulfill two criteria:
\begin{enumerate}
\item In the bulk isotropic and nematic phase, the influence of the 
constraint must vanish. In other words, $E_c$ may not affect the bulk 
properties of the unconstrained system.
\item In the coexistence region, where the system schematically resembles 
\fig{snap}(b), the director $\vec{n}$ of the nematic phase must point 
along some specified tilt angle $\theta$.
\end{enumerate}
As it turns out, a suitable constraint can be written as:
\begin{equation}\label{eq:con}
  E_c =
  \begin{cases}
  0 & |90 - \arccos |\vec{n} \cdot \vec{z}| - \theta_t| < \delta, \\
  \infty & {\rm otherwise}.
  \end{cases}
\end{equation}
Here, $\vec{n}$ is the nematic director, defined in \sect{op}. The angles 
$\theta_t$ and $\delta$ are inputs of the method, and must be specified 
beforehand. By using the constraint, only states whose angle between 
director and $xy$-plane is within $\theta_t \pm \delta$ are retained, 
while all other states are rejected.

For large systems, \eq{eq:con} does not affect bulk properties, since bulk 
properties are insensitive to the overall orientation of the phase. In 
contrast, in the coexistence region, the constraint has a dramatic effect. 
In these cases, approximately half of the simulation box is filled with an 
isotropic domain, and the other half with a nematic domain, see 
\fig{snap}(b). Due to the constraint, the angle between the director of 
the nematic domain and the $xy$-plane is within $\theta_t \pm \delta$. At 
the same time, the use of an elongated simulation box forces the 
interfaces to form in the $xy$-plane as well. In other words, by setting 
$\theta_t$, the anchoring angle $\theta$ can be fixed. More precisely, one 
has $\theta = \theta_t$. Naturally, the threshold angle $\delta$ should be 
chosen as small as possible, while, at the same time, maintaining 
reasonable simulational efficiency. The optimal value is model dependent, 
and best obtained using trial-and-error.

The idea to obtain $\gt$ is now clear. We first choose a tilt angle 
$\theta$ of interest. Next, we measure the order parameter distribution 
$P(X)$, in an elongated simulation box, using the methodology of 
\sect{opd}. In addition, we incorporate the constraint of \eq{eq:con} in 
the simulations, using $\theta_t = \theta$. The peak positions in $P(X)$ 
should again yield the bulk properties of the coexisting isotropic and 
nematic phase. The barrier $\Delta F$, see \fig{bimodal}(b), can be 
plugged into \eq{eq:st} to obtain the interfacial tension at the chosen 
anchoring angle $\theta$. Since bulk properties should not be affected by 
the constraint, we expect the peak positions in $P(X)$ to coincide with 
those of an unconstrained simulation. In contrast to the unconstrained 
simulations, a dependence of the interfacial tension on the anchoring 
angle $\theta$ is anticipated. To what extent these expectations are met 
in actual simulations will be investigated next.

\section{Results: lattice simulations}
\label{res1}

\subsection{Lattice model and motivation}

As announced in the Introduction, we first test our method in a lattice 
model of a thermotropic liquid crystal. The aim is to measure 
$\gt$. The simulations are performed on a three-dimensional 
periodic lattice of size $V = L \times L \times D$, with $D \gg L$. To 
each lattice site $i$, a liquid crystal is attached with (normalized) 
orientation $\vec{d}_i$. The liquid crystals interact via the potential
\begin{equation}\label{eq:ll}
 E =  - \epsilon
 \sum_{\langle i,j \rangle} \sigma_{ij} | \vec{d}_i \cdot \vec{d}_j |^p, 
\end{equation}
where the summation is over nearest neighbors, coupling constant 
$\epsilon$, and exponent $p>0$. In what follows, factors of $\kb T$ are 
absorbed in the coupling constant $\epsilon$, with $T$ the temperature, 
and $\kb$ the Boltzmann constant. The ``anisotropy'' parameter is given by
\begin{equation}\label{eq:ani}
  \sigma_{ij} = 1 + \nu \left[ (\vec{d}_i \cdot \vec{r}_{ij})^2
  + (\vec{d}_j \cdot \vec{r}_{ij})^2 \right],
\end{equation}
with $\vec{r}_{ij}$ a unit vector pointing from site $i$ to $j$, and $\nu$ 
a parameter between $-0.5 \leq \nu \leq 0.5$. On a cubical lattice, each 
site has six nearest neighbors. Consequently, there are only three 
possible axes along which the vectors $\vec{r}_{ij}$ can be oriented.

For $p=2$ and $\nu=0$, this model reduces exactly to the LL model 
\cite{physreva.6.426}. In this case, the model exhibits a first-order IN 
transition, but it is very weak \cite{physrevlett.69.2803}. This makes the 
LL model rather inconvenient for our purposes. For example, to stabilize 
two interfaces so as to recover the coexistence of \fig{snap}(b), huge 
systems would be required. Such large-scale simulations are not the aim of 
the present work, and so we have chosen to modify the interactions 
appropriately. More precisely, we use a larger exponent in \eq{eq:ll}, 
namely $p=10$. The effect of this is a sharper pair interaction, meaning 
that neighboring molecules only lower their energy when they are closely 
aligned. It is known that, under such interactions, first-order phase 
transitions become enhanced \cite{physrevlett.89.285702, 
physrevlett.52.1535, physrevlett.88.047203, enter.romano.ea:2006, 
vink:2006*b} (even in two dimensions).

\begin{figure}
\begin{center}
\includegraphics[clip=,width=0.65\columnwidth]{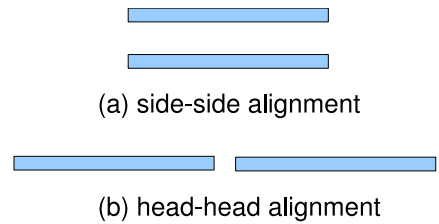}
   
\caption{\label{ani} Illustration of the spatial anisotropy in the liquid 
crystal pair interaction. In a realistic system, the energies of the above 
two configurations will generally differ. The LL model, however, makes no 
distinction.}

\end{center}
\end{figure}

By using $p=10$ in \eq{eq:ll}, the model is expected to exhibit a strong
first-order IN transition. Nevertheless, this is not sufficient to study
anchoring because, for $\nu=0$, the interactions are spatially isotropic. 
In other words, the interactions do not depend on the relative positions
of the molecules, and so they cannot produce any anchoring effects at the
IN interface \cite{physreve.68.041709}. In realistic systems, the particle
interactions are typically anisotropic, see \fig{ani}. Shown are two
liquid crystal arrangements, labeled (a) and (b). Even though the
orientations of the molecules are identical in both cases, it is clear
that the energies need not be the same. In the LL model, however, for
which $\nu=0$, there is no distinction between the two arrangements. In
order to nevertheless study anchoring effects, we allow $\nu \neq 0$ in
\eq{eq:ani}, in which case the model does make the distinction. More
precisely, we have $\sigma_{ij} = 1$ for case (a), and $\sigma_{ij} =
1+2\nu$ for case (b). By choosing $\nu<0$, side-side alignment is
energetically favored; choosing $\nu>0$ favors head-head alignment. For
$\nu=0$, the interactions are isotropic, in which case no particular
alignment is preferred.

\subsection{Bulk phase behavior}

We first determine the bulk behavior of \eq{eq:ll}. Recall that we keep 
the exponent fixed at $p=10$. The aim is to measure the variation of the 
bulk properties as a function of $\nu$. More precisely, we consider the 
transition inverse temperature $\et$, and the coexistence energy densities 
of the isotropic and nematic phase. To this end, we use the simulation 
methodology of \sect{opd} {\it without} the constraint of \eq{eq:con}. The 
energy distribution $P(E)$ is measured in a MC simulation, using a biased 
sampling scheme \cite{virnau.muller:2004}, at $\epsilon=0$. Histogram 
reweighting \cite{ferrenberg.swendsen:1988} is used to determine the 
value of $\epsilon$ for which the ``equal-area'' rule is obeyed, yielding 
$\et$. The energy densities are then read-off from the peak positions. An 
example distribution $P(E)$ is shown in \fig{bimodal}. The simulations are 
performed using single particle MC moves, whereby a random orientation is 
assigned to a randomly selected lattice site, accepted with the Metropolis 
criterion \cite{newman.barkema:1999}. Typical lattice sizes are $L = 
10 - 20$ and $D = 20 - 40$. The CPU time required to obtain $P(E)$ 
accurately for a large system is around 48 hours.

\begin{figure}
\begin{center}
\includegraphics[clip=,width=\figwidth]{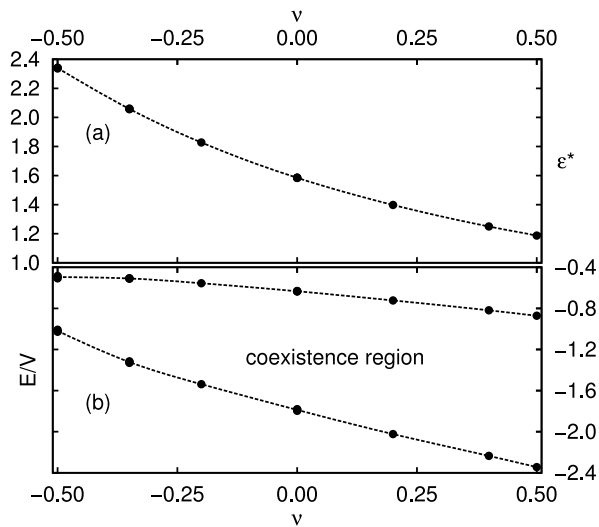}
\caption{\label{phase} Bulk properties of \eq{eq:ll} with $p=10$ as a 
function of $\nu$. Points are actual simulation data; curves serve to 
guide the eye. (a) Variation of $\et$ with $\nu$. (b) Binodal curves, 
showing the energy density $E/V$ of the isotropic phase (top curve), and 
of the nematic phase (lower curve), as a function of $\nu$. At energy 
densities between the curves, coexistence between isotropic and nematic 
domains occurs; simulation snapshots will then schematically resemble 
\fig{snap}(b).}
\end{center}
\end{figure}

The variation of $\et$ with $\nu$ is shown in \fig{phase}(a). The behavior 
is simply monotonic: by increasing $\nu$, $\et$ goes down. The energy 
densities, shown in \fig{phase}(b), reveal more interesting behavior. By 
decreasing $\nu$, the energy difference between the isotropic and the 
nematic phase becomes smaller. In other words, the transition becomes 
weaker. The simulations do not rule out that the curves meet when $\nu$ 
becomes sufficiently negative, possibly terminating in a critical point, 
but clearly additional efforts are required to resolve this. All that 
matters for the present work, however, is the fact that \fig{phase}(b) 
reveals a large coexistence region, over a substantial range of $\nu$ 
values. This confirms our expectation that, by using $p=10$ in \eq{eq:ll}, 
the first-order nature of the transition is enhanced significantly. This 
makes the model ideal to study anchoring effects, with which we proceed 
next.

\subsection{Anchoring effects for $\nu=0$}

As a benchmark, we consider \eq{eq:ll} with $\nu=0$, using the newly 
proposed method. Note that, for $\nu=0$, the model is spatially isotropic, 
and so we do not expect to see any anchoring effects. We MC simulate 
\eq{eq:ll} as before, with the constraint of \eq{eq:con} explicitly 
included. Two anchoring conditions are considered: homogeneous and 
homeotropic. Recall that the anchoring is set via $\theta_t$ in 
\eq{eq:con}. For the threshold angle, we use $\delta=0.75$~degrees. In 
addition to $P(E)$, we also measure $P(S)$, with $S$ the nematic order 
parameter defined in \sect{op}. For completeness, we mention that our 
simulations are performed using a bias on the nematic order parameter $S$, 
see details in \olcite{vink.wolfsheimer.ea:2005}.

\begin{figure}
\begin{center}
\includegraphics[clip=,width=\figwidth]{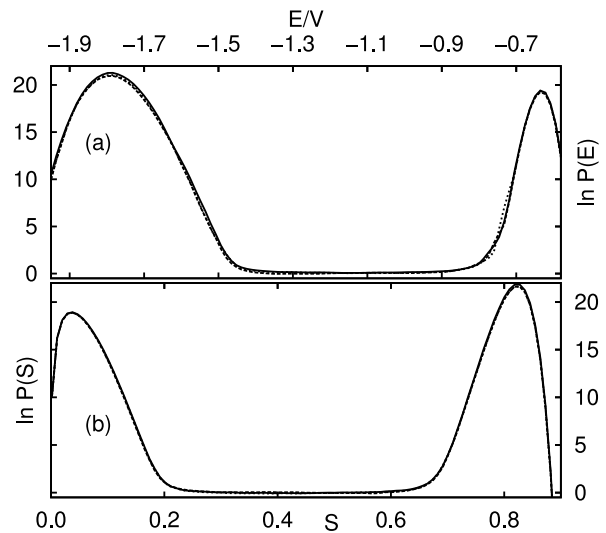}

\caption{\label{nu_zero} Order parameter distributions of \eq{eq:ll} with 
$p=10$ and $\nu=0$, obtained using box dimensions $L=10$ and $D=30$. Shown 
are the logarithm of the energy distribution (a), and of the nematic order 
parameter (b). In each case, three distributions are shown, corresponding 
to homogeneous and homeotropic enforced anchoring, as well as no enforced 
anchoring direction. The curves overlap almost perfectly, indicating the 
absence of any anchoring effects.}

\end{center}
\end{figure}

The resulting energy distributions are given in \fig{nu_zero}(a), which 
actually shows three distributions. Shown are the two distributions 
obtained using the new method, corresponding to homogeneous and 
homeotropic anchoring, as well as the distribution obtained without any 
enforced anchoring. The striking feature is that the curves overlap almost 
perfectly. This result is crucial because it demonstrates the consistency 
of the method. First of all, the insensitivity of the peak positions with 
respect to the enforced anchoring, confirms that {\it bulk} properties are 
{\it not} affected by the constraint. In addition, we find that the 
barrier $\Delta F$, defined in \fig{bimodal}(b), also does not depend on 
the anchoring condition. In other words, the interfacial tension is 
independent of the tilt angle, which is precisely what one expects for an 
isotropic potential. Additional confirmation of the consistency of the new 
method is provided in \fig{nu_zero}(b), which shows the corresponding 
distributions $\ln P(S)$ of the nematic order parameter. Again, the curves 
overlap almost perfectly. Note also that, for the interfacial tension, it 
does not matter whether one reads-off the barrier height in $\ln P(E)$ or 
$\ln P(S)$. As \fig{nu_zero} shows, the barriers are nearly equal (the 
slight variation gives an indication of the statistical uncertainty).

\begin{figure}
\begin{center}
\includegraphics[clip=,width=\figwidth]{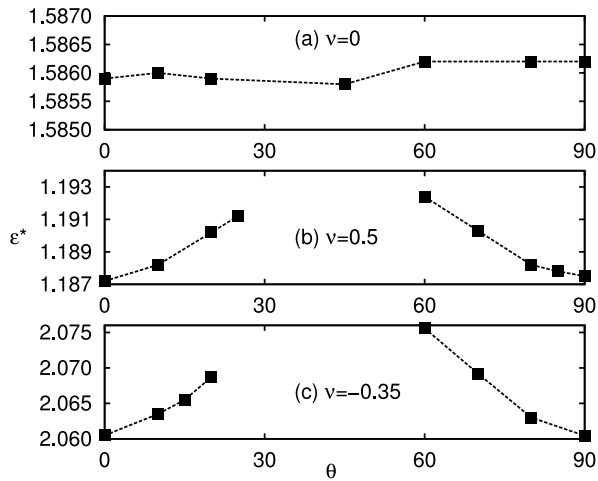}

\caption{\label{et} Transition inverse temperature $\et$ versus anchoring 
angle $\theta$ for \eq{eq:ll} with $p=10$ and three values of $\nu$ as 
indicated. Closed symbols are actual simulation data; lines serve to 
guide the eye. The data were obtained using box dimensions $L=15$ and 
$D=40$. Note the much finer scale in (a) compared to (b) and (c).}

\end{center}
\end{figure}

\begin{figure}
\begin{center}
\includegraphics[clip=,width=\figwidth]{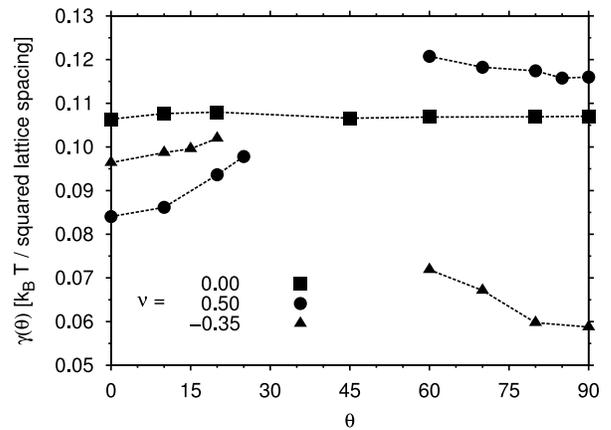}

\caption{\label{gam} Angle dependent interfacial tension $\gt$ 
of \eq{eq:ll} using $p=10$ and three values of $\nu$ as indicated. Closed 
symbols are actual simulation data; lines serve to guide the eye. The 
data were obtained using box dimensions $L=15$ and $D=40$.}

\end{center}
\end{figure}

We have repeated the above analysis using larger lattices, considering 
also tilt angles between 0 and 90 degrees. Shown in \fig{et}(a) is the 
transition inverse temperature $\et$ versus $\theta$. As expected, for the 
spatially isotropic case, $\et$ is insensitive to $\theta$, and we obtain 
$\et = 1.5860 \pm 0.0005$. Shown in \fig{gam} is the angle dependent 
interfacial tension $\gt$, as extracted from the barrier 
$\Delta F$ in $\ln P(S)$ and using \eq{eq:st}. Here, $\Delta F$ was taken 
to be the average height of the peaks, measured with respect to the flat 
region. As expected, the interfacial tension does not display any 
pronounced $\theta$ dependence (the variation stays below 2\%). For 
$\nu=0$, we thus find $\gamma = 0.108 \pm 0.002 \, \kb T$ per squared 
lattice spacing, independent of the tilt angle.

\subsection{Anchoring effects for $\nu=0.5$}

\begin{figure}
\begin{center}
\includegraphics[clip=,width=\figwidth]{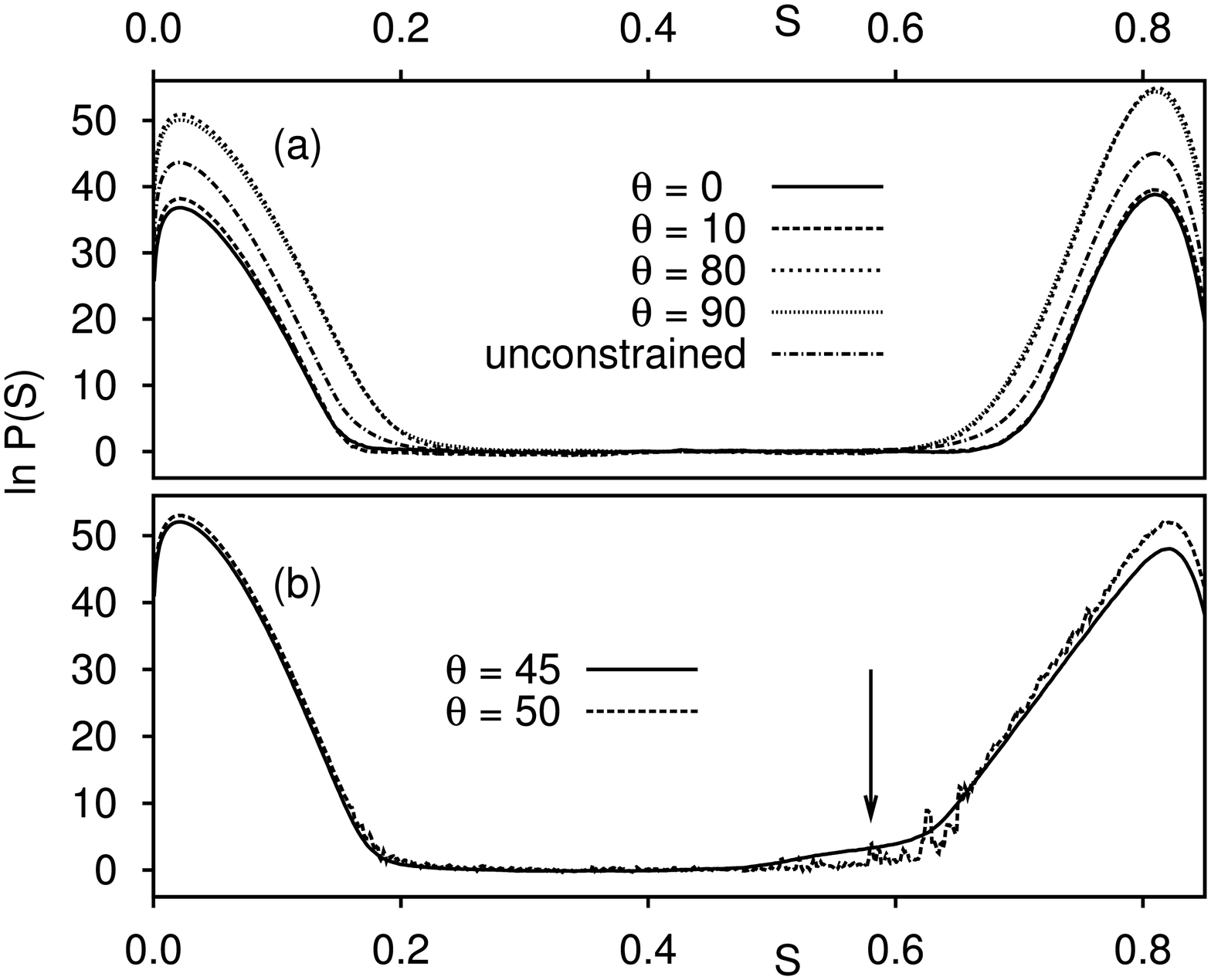}

\caption{\label{nu_pos} Logarithm of the nematic order parameter 
distribution $P(S)$, at coexistence, of \eq{eq:ll} with $p=10$ and 
$\nu=0.5$. Shown is $\ln P(S)$ for various imposed anchoring angles 
$\theta$, with $\theta$ the angle between the nematic director and the 
plane of the IN interface. The distributions were obtained using box 
dimensions $L=15$ and $D=40$.}

\end{center}
\end{figure}

\begin{figure*}
\begin{center}
\includegraphics[clip=,width=2.1\columnwidth]{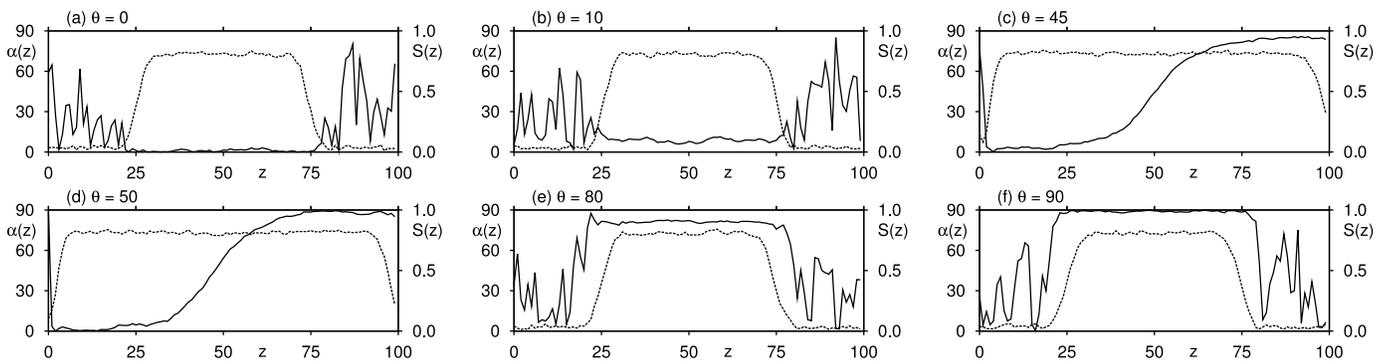}

\caption{\label{prof} Profiles $S(z)$ (dashed curves) and $\alpha(z)$ 
(solid curves) of \eq{eq:ll} with $p=10$ and $\nu=0.5$, for various 
imposed tilt angles $\theta$. The profiles were obtained at the transition 
inverse temperature $\et$, overall nematic order parameter $S=0.4$, and 
box dimensions $L=40$ and $D=100$.}

\end{center}
\end{figure*}

Having verified that the spatially isotropic case $\nu=0$ does not reveal 
any anchoring effects, we now consider \eq{eq:ll} using $\nu=0.5$. In this 
case, the model becomes anisotropic, and the tilt angle of the nematic 
phase with respect to the IN interface should become a relevant parameter. 
For a number of tilt angles, we have measured the order parameter 
distribution $\ln P(S)$ at coexistence; recall that the tilt angle is set 
via the constraint of \eq{eq:con}. Typical distributions are plotted in 
\fig{nu_pos}. Shown in (a) are distributions for tilt angles close to 
homogeneous and homeotropic anchoring; shown in (b) are distributions for 
two ``in-between'' tilt angles. Also shown in (a) is the ``unconstrained'' 
distribution, which one obtains without imposing the constraint of 
\eq{eq:con}. For tilt angles that are close to $\theta=0$ or $\theta=90$, 
the distributions behave as expected: they are bimodal, and also exhibit a 
pronounced flat region between the peaks. In addition, we observe that the 
barrier height, defined in \fig{bimodal}, depends profoundly on the 
imposed tilt angle. Since the barrier is related to the interfacial 
tension, via \eq{eq:st}, we can already see that anchoring effects are 
present. Interestingly, for the ``in-between'' tilt angles, bimodal 
distributions can also be identified, but the region between the peaks is 
not quite flat, see the arrow in \fig{nu_pos}(b). This suggests that, for 
these ``in-between'' angles, the constraint does not quite produce the IN 
coexistence scenario of \fig{snap}, but rather something else.

To verify what is going on, we have generated a number of snapshots, at 
{\it fixed} nematic order parameter $S=0.4$. For all distributions in 
\fig{nu_pos}, this value is well between the peak positions. The snapshots 
are generated at the transition inverse temperature $\et$ using MC 
simulation. After equilibration, we collect the profiles $S(z)$ and 
$\alpha(z)$. Here, $S(z)$ is the nematic order parameter in the $z$-th $L 
\times L$ slab perpendicular to the elongated $\hat{z}$-direction, and 
$\alpha(z)$ the angle between the director in that slab and the 
$xy$-plane. The profiles are shown in \fig{prof}, for the same tilt angles 
as studied in \fig{nu_pos}. For $\theta=0,10,80,90$~degrees, the profile 
$S(z)$ strikingly confirms IN phase coexistence. We can clearly identify 
one region where $S(z)$ is close to zero, corresponding to the isotropic 
phase, and another region where $S(z)$ is closer to unity, corresponding 
to the nematic phase. Moreover, in the nematic phase, $\alpha(z)$ is 
roughly constant, and the plateau value closely follows the imposed 
anchoring angle $\theta$. In other words, the constraint has the expected 
effect, namely to force the nematic phase to assume a specified tilt 
angle. Of course, in the isotropic phase, there is no preferred direction, 
and $\alpha(z)$ fluctuates randomly; one can show that the average should 
converge to $90 (\pi-2)/\pi \approx 32.7$ degrees. In contrast, for 
$\theta=45,50$~degrees, the scenario is completely different. Here, $S(z)$ 
is roughly constant at $S \approx 0.8$, implying a single nematic phase 
along the entire $\hat{z}$-direction. In addition, from the corresponding 
$\alpha(z)$, we see that the nematic is twisted: starting at $z=0$, 
$\alpha(z)$ rotates smoothly from 0 to 90 degrees, abruptly dropping back 
to 0 again as one passes through the periodic boundary at $z=100$. 
Clearly, this configuration does not reflect IN coexistence at all, but 
rather a twisted nematic phase with a surface defect.

In light of \fig{prof}, it is clear that the free energy barrier for 
``in-between'' tilt angles does not reflect the interfacial tension, and 
consequently \eq{eq:st} does not apply. For angles that are close to 
homogeneous and homeotropic anchoring, however, the IN scenario of 
\fig{snap} is confirmed. Therefore, for these angles, we may use 
\eq{eq:st} to obtain the angle dependent interfacial tension 
$\gt$. The result is shown in \fig{gam}, which reveals several 
trends. First of all, in contrast to $\nu=0$, we now observe a profound 
variation of $\gt$ with the imposed tilt angle. The interfacial 
tension is smallest at $\theta=0$, corresponding to homogeneous anchoring. 
We therefore expect unconstrained simulations, whereby $\theta$ is not 
imposed but freely fluctuating, to mostly exhibit homogeneous anchoring. 
However, the data of \fig{gam} also suggest the presence of a shallow 
minimum at $\theta=90$, which corresponds to homeotropic anchoring. In 
other words, for $\nu=0.5$, homeotropic anchoring appears to be 
metastable. Since the difference in interfacial tension between 
homogeneous and homeotropic anchoring is small, it is not {\it a-priori} 
clear which anchoring condition will actually prevail in an unconstrained 
simulation. 

We have therefore performed a number of unconstrained simulations, 
i.e.~without \eq{eq:con}, and measured the coexistence distribution $\ln 
P(S)$. In addition, for each simulation, we also recorded the $n_z$ 
component of the director $\vec{n}$ as a function of $S$. In some cases, 
we found that the system selects $\theta=0$, in which case $n_z$ drops to 
zero once nematic order sets in, but quite often also $\theta=90$ is 
selected, in which case $n_z$ becomes close to unity. More precisely, 
using lattice dimensions $L=15$ and $D=40$, we performed 90~unconstrained 
simulations and found that metastable homeotropic anchoring ($\theta=90$) 
was selected 27~times, i.e.~in 30\% of the cases. This finding is 
important because it shows that the order parameter distribution of the 
unconstrained simulation actually reflects a ``weighted average'' of both 
stable and metastable anchoring. This feature is illustrated in 
\fig{nu_pos}(a), which also includes $\ln P(S)$ of the unconstrained 
simulation. As the figure shows, the free energy barrier of the 
unconstrained simulation is somewhere ``in-between'' homogeneous and 
homeotropic anchoring.

Another important finding is that the unconstrained simulations reveal 
only stable anchoring $(\theta=0)$, and metastable anchoring 
$(\theta=90)$, while no other anchoring angles were observed. This result 
is consistent with $\gt$ of \fig{gam}, which indeed features 
just two minima. In other words, all ``in-between'' tilt angles are 
unstable. Systems in which the anchoring is held artificially fixed at 
such unstable angles, for example via the constraint of \eq{eq:con}, will 
experience an additional strain. For highly unstable tilt angles, the 
strain is so strong, that it becomes favorable for the system to break-up 
the IN interfaces altogether, and form a twisted nematic. This is 
precisely the effect we observed for $\theta=45,50$ in \fig{prof}. 
However, also for tilt angles close to the stable and metastable angle, we 
noticed that the strain manifests itself. In this case, a small shift in 
the transition inverse temperature $\et$ can be detected. The effect is 
illustrated in \fig{et}(b), which shows $\et$ as a function of the imposed 
tilt angle $\theta$. For unstable tilt angles, $\et$ is systematically 
larger compared to the stable and metastable angles. Of course, for the 
stable and metastable angles, which are the experimentally relevant cases, 
one finds the same transition temperature again. Note also \fig{et}(a), 
which shows that the effect for the spatially isotropic potential $\nu=0$ 
does not occur, as expected.

\begin{figure}
\begin{center}
\includegraphics[clip=,width=\figwidth]{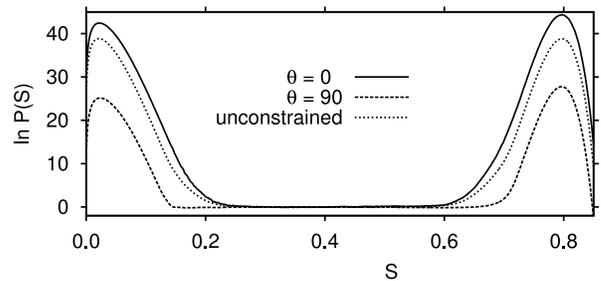}
\caption{\label{nu_neg} Coexistence distributions $\ln P(S)$ of 
\eq{eq:ll}, with $p=10$ and $\nu = -0.35$, using box dimensions $L=15$ and 
$D=40$. Shown are distributions for imposed tilt angles $\theta=0,90$ 
degrees, as well as the unconstrained distribution that one obtains when 
the tilt angle is allowed to freely fluctuate.}
\end{center}
\end{figure}

\subsection{Anchoring effects for $\nu=-0.35$}

For completeness, we also performed a number of simulations using a 
negative value of $\nu$ in \eq{eq:ll}, namely $\nu = -0.35$. Recall that 
for negative values, the side-side arrangement of \fig{ani} becomes 
energetically more favorable. Compared to $\nu=0.5$, one might intuitively 
expect that this reverses the stable and metastable anchoring angles. The 
angle dependent interfacial tension indeed confirms this, see \fig{gam}. 
We now observe that homeotropic anchoring yields the lowest interfacial 
tension, i.e.~is stable, while homogeneous anchoring appears to be 
metastable. In agreement with $\nu=0.5$, we again measure a shift in $\et$ 
when unstable anchoring angles are imposed, see \fig{et}(c). 
Interestingly, even though for $\nu=-0.35$ homeotropic anchoring yields 
the lowest interfacial tension, we observed that {\it unconstrained} 
simulations have difficulty ``finding'' this configuration. During a 
series of 95 unconstrained simulation runs, homeotropic anchoring was 
selected only 35~times, i.e.~in 37\% of the cases. In other words, the 
interfacial tension extracted from $\ln P(S)$ in the unconstrained 
simulation, rather reflects the metastable anchoring condition, see 
\fig{nu_neg}. The figure clearly shows that homeotropic anchoring 
($\theta=90$) yields the lowest free energy barrier, while the barrier in 
the unconstrained distribution is significantly higher (and, in fact, 
rather closely resembles homogeneous anchoring). From a computational 
point of view, the result of \fig{nu_neg} is important because it shows 
that simulations do not generally find the optimal anchoring angle by 
themselves, even in a relatively simple lattice model.

\section{Results: soft rods}
\label{res2}

\begin{figure}
\begin{center}
\includegraphics[clip=,width=\figwidth]{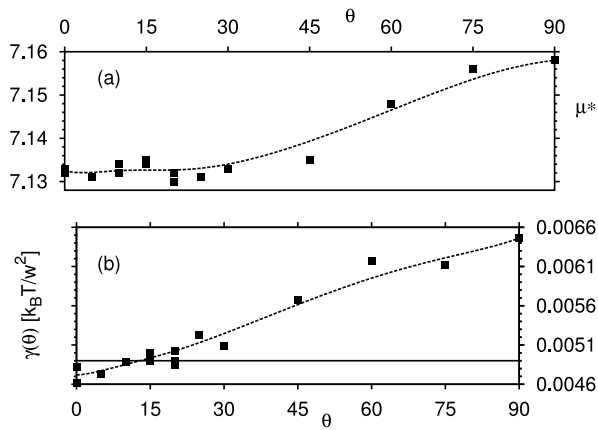}

\caption{\label{rods} Anchoring properties of soft rods at the IN 
transition. Shown in (a) is the coexistence chemical potential $\mt$ 
versus $\theta$; in (b) the angle dependent interfacial tension $\gt$ 
versus $\theta$. Closed squares are raw simulation data; the curves serve 
to guide the eye. The horizontal line in (b) marks the interfacial tension 
of an {\it unconstrained} simulation, taken from previous work 
\cite{vink.wolfsheimer.ea:2005}. The simulations were performed using box 
dimensions $L=35$ and $D=105$.}

\end{center}
\end{figure}

Next, we investigate anchoring effects in an {\it off-lattice} fluid of 
soft rods. The rods are modeled as spherocylinders, of length $l$ and 
width $w$. In this section, we set $l/w=10$, and $w$ will be the unit of 
length. The rods interact via a repulsive pair potential, whereby rod 
overlap is penalized with an energy cost of $2 \, \kb T$. For more details 
about the model, the reader is referred to previous work 
\cite{vink.schilling:2005, vink.wolfsheimer.ea:2005}. The rods are 
simulated in the grand-canonical ensemble, i.e.~at constant temperature 
$T$, chemical potential $\mu$, and system volume $V$, while the number of 
rods in the system fluctuates. Again, we use an elongated simulation box 
$V = L \times L \times D$, with periodic boundary conditions. The 
simulations are performed using standard insertion/deletion moves 
\cite{frenkel.smit:2001}, and the distribution $\ln P(S)$ is recorded, 
defined as the probability to observe the nematic order parameter $S$, at 
the specified tilt angle $\theta$. As before, $\theta$ is imposed using 
the constraint of \eq{eq:con}. For soft rods, we noticed that a 
substantially larger threshold angle was needed to maintain efficiency. 
Here, we used $\delta=2.5$~degrees. Whereas in the thermotropic liquid 
crystal of \eq{eq:ll} phase coexistence is achieved by tuning the inverse 
temperature $\epsilon$, here that role is played by the chemical potential 
$\mu$. At the coexistence chemical potential $\mt$, $\ln P(S)$ 
becomes bimodal: coexistence properties and interfacial tensions may then 
be extracted from the peak positions and heights, as in \fig{bimodal}.

The results of the soft rod simulations are summarized in \fig{rods}. 
Compared to the lattice simulations of \eq{eq:ll}, the data reveal 
significant scatter. This indicates that soft rod simulations are 
demanding, and already close to the limit of what is currently tractable. 
Nevertheless, a number of trends emerge. According to \fig{rods}(b), $\gt$ 
increases monotonically with $\theta$, with the minimum occurring at 
$\theta=0$. In other words, soft rods favor homogeneous anchoring, and the 
presence of metastable angles is unlikely. The data also show that the 
anchoring angle is a remarkably ``soft'' degree of freedom: the free 
energy cost of tilting the nematic director away from the IN interface is 
small. This is apparent from the coexistence chemical potential, see 
\fig{rods}(a). Note that \fig{rods}(a) is the ``analogue'' of \fig{et} for 
the lattice model of \eq{eq:ll}. For the lattice model, the coexistence 
inverse temperature increases profoundly away from the stable and 
metastable angles. This increase is a manifestation of the strain 
introduced into the system when {\it unstable} anchoring angles are 
imposed. In contrast, for soft rods, the coexistence chemical potential 
remains nearly constant over a wide range; only when $\theta>30$ or so, 
does $\mt$ begin to exhibit a pronounced $\theta$ dependence. For soft 
rods, the anchoring angle can thus be varied around the stable direction 
over a fairly large range, without introducing excessive strain into the 
system. This result is important for {\it unconstrained} simulations, 
where the anchoring angle is allowed to fluctuate freely. It is unlikely 
that such simulations would always reveal homogeneous anchoring. Rather, 
we expect a range of anchoring angles $0 < \theta < 30$ to be present. The 
horizontal line in \fig{rods}(b) marks the interfacial tension obtained 
during an {\it unconstrained} simulation of soft rods 
\cite{vink.wolfsheimer.ea:2005}, and indeed confirms this expectation. 
Even though the lowest interfacial tension is obtained at $\theta=0$, the 
unconstrained simulation slightly exceeds this value. Instead, it rather 
reflects the average of $\gt$ over the range $0 < \theta < 30$~degrees. 
Additional confirmation is obtained from simulation snapshots of {\it 
unconstrained} simulations, which reveal substantial fluctuations of the 
anchoring angle around the homogeneous direction.

\section{Summary and outlook}

In this paper, an alternative simulation approach to study anchoring 
effects at the IN interface in liquid crystals was described. In 
particular, we focused on the angle dependent interfacial tension $\gt$, 
with $\theta$ the anchoring or tilt angle. The proposed method is based on 
recent innovations \cite{vink.schilling:2005, vink.wolfsheimer.ea:2005} 
where the order parameter distribution is used to extract interfacial 
properties. The new twist has been to introduce a constraint into the 
Hamiltonian, see \eq{eq:con}, which forces the nematic director to 
maintain a specified angle with respect to the $xy$-plane. The idea is 
that, by using a simulation box that is elongated in the $z$-direction, IN 
interfaces will form in the $xy$-plane as well. The constraint then allows 
the anchoring angle $\theta$ to be fixed to some value of interest.

At the same time, a new liquid crystal model was introduced. The model is 
defined on a lattice and exhibits a strong first-order IN transition. In 
addition, the preferred anchoring (homogeneous, homeotropic, or neutral) 
can be tuned by means of a single parameter. Compared to more elaborate 
{\it off-lattice} models, such as rods or platelets, the lattice variant 
is considerably easier to simulate. In particular, equilibration is less 
problematic, and high-quality data are readily generated. Precisely this 
property was exploited to obtain $\gt$ for the lattice model, using the 
new method. Indeed, when anchoring effects are ``switched-off'', by 
setting $\nu=0$ in \eq{eq:ani}, $\gt$ becomes constant. In contrast, when 
$\nu \neq 0$, a pronounced $\theta$ dependence is revealed. For these 
cases, only homogeneous and homeotropic anchoring were seen to be 
relevant. More precisely, for $\nu>0$, homogeneous anchoring is stable, 
and homeotropic anchoring metastable. For $\nu<0$, the trend is reversed. 
In other words, the preferred anchoring depends sensitively on the details 
of the interactions. Our results have also shown that, when unstable 
anchoring angles are imposed, the new method must be used with some care. 
In those cases, the simulations do not reveal IN coexistence, but rather a 
twisted nematic phase. Fortunately, when this happens, the method gives a 
clear warning, in the form of a shift in the coexistence temperature. A 
somewhat surprising finding was that, even for the simple lattice model, 
simulations do not generally find the ``optimal'' anchoring angle by 
themselves. Instead, when the nematic director is allowed to fluctuate 
freely, both stable and metastable anchoring are typically revealed.

We have also applied the new method to obtain $\gt$ for a fluid of soft 
rods. For soft rods, anchoring effects could also be identified, albeit 
that the data are significantly less accurate. The simulations reveal 
homogeneous anchoring to be stable, a finding which is consistent with 
most theoretical studies of hard rods. Interestingly, for soft rods, no 
metastable anchoring angle could be detected, which makes this model 
qualitatively very different from the lattice model of \eq{eq:ll}. It 
confirms, once again, that anchoring effects are extremely sensitive to 
the particle interactions.

For the future, investigations of the capillary wave spectrum for the 
lattice model of \eq{eq:ll} are planned. As mentioned in the Introduction, 
the spectrum is qualitatively affected by the anchoring condition 
\cite{akino.schmid.ea:2001, elgeti.schmid:2005, schmid.germano.ea:2007}. 
Since, in \eq{eq:ll}, the anchoring can be tuned using a single parameter, 
and since the model is easy to simulate anyhow, such investigations should 
be worthwhile. A sound understanding of the lattice model may well be a 
prerequisite before more complicated {\it off-lattice} simulations are 
attempted.

\acknowledgments

This work was supported by the {\it Deutsche Forschungsgemeinschaft} under 
the SFB-TR6 (project section D3).

\bibstyle{revtex} 
\bibliography{mainz}

\end{document}